\pdfoutput=1
\documentclass[a4paper,12pt]{article}
\usepackage{graphicx}
\usepackage{amssymb}
\usepackage{amsmath}
\usepackage{amsbsy}
\usepackage{color}
\usepackage{dsfont}
\usepackage{upgreek}
\usepackage{soul}
\usepackage{cite}
\usepackage[normalem]{ulem}
\newcommand{\ee}{\end{equation}}
\newcommand{\bb}{\begin{equation}}
\newcommand{\be}{\begin{equation}}
\newcommand{\eqb}{\begin{eqnarray}}
\newcommand{\eqf}{\end{eqnarray}}

\definecolor{gre}{rgb}{0,0.4,0.3}
\usepackage{chngpage}

\usepackage{color}

\def\bb{\begin{equation}}
\def\ee{\end{equation}}
\def\ba{\begin{eqnarray}}
\def\ea{\end{eqnarray}}
\def \eqf{\begin{eqnarray}}
\def \eea{\end{eqnarray}}
\def \eqb{\begin{eqnarray}}
\def \eqf{\end{eqnarray}}

\usepackage{graphics}
\usepackage{amsmath,amssymb}
\usepackage{graphicx}
\usepackage{dcolumn}
\usepackage{bm}
\usepackage{epsfig}
\usepackage{graphicx}
\usepackage{dcolumn}
\usepackage{graphicx,epsfig}%
\def \ee{\end{equation}}
\def \be{\begin{equation}}
\def \bea{\begin{eqnarray}}
\def \eea{\end{eqnarray}}

\begin{document}

\title{
Hidden Photons in Aharonov-Bohm-Type Experiments}
\author{
Paola Arias$^1$,
Christian Diaz$^2$,
Marco Aurelio Diaz$^2$, 
Joerg Jaeckel$^3$,\\
Benjamin Koch$^2$ and  
Javier Redondo$^{4,5}$\\[2ex] 
\small{\em $^1$Departmento de F\'isica, Universidad de Santiago de Chile,} 
\small{\em Casilla 307, Santiago, Chile} \\[0.5ex]  
\small{\em $^2$Instituto de F\'{i}sica,
Pontificia Universidad Cat\'{o}lica de Chile,} \\ 
\small{\em Av. Vicu\~{n}a Mackenna 4860, 
Santiago, Chile }\\[0.5ex]  
\small{\em $^3$Institut f\"ur Theoretische Physik, Universit\"at Heidelberg,} \\
\small{\em Philosophenweg 16, 69120 Heidelberg, Germany}\\[0.5ex]
\small{\em $^4$University of Zaragoza, P. Cerbuna 12, 50009 Zaragoza, Spain} \\[0.5ex]  
\small{\em $^5$Max Planck Institut f\"ur Physik, F\"ohringer Ring 6, 80803 M\"unchen, Germany} \\[0.5ex]  
}
\maketitle

\begin{abstract}
We discuss the Aharonov-Bohm effect in the presence of hidden photons kinetically mixed with the ordinary electromagnetic photons. The hidden photon field causes a slight phase shift in the observable interference pattern. 
It is then shown how the limited sensitivity  of this experiment can be largely
improved. 
The key observation is that the hidden photon field causes a leakage of the ordinary magnetic field into the supposedly field-free region. The direct measurement of this magnetic field can provide a sensitive experiment with a good discovery potential, particularly below the $\sim$ meV mass range for hidden photons.
\end{abstract}


\maketitle
\newpage
\section{Introduction}
Astrophysical and cosmological observations give clear evidence that 95\% of the Universe is made 
out of substances not represented in the Standard Model. 
Yet, again and again the Standard Model persists in experimental tests and new physics remains elusive.

One possible explanation, of why new physics is so elusive is that it could reside in a so-called hidden sector, 
that couples only very weakly with Standard Model particles and therefore with our experiments. 
In this case new particles do not need to be heavy to evade detection.
Thus, they would not be found by experimental efforts that concentrate on higher and higher energies.
Instead, to probe such hidden sectors one needs new, very precise experiments.

Besides the purely phenomenological argument that hidden sectors are a good way to ``hide'' the new physics, theoretical model building also provides motivation for their existence. Indeed, hidden sectors are a common feature of many proposed extensions to the Standard Model. 
One of the simplest versions of a hidden sector is an extra U(1) gauge degree of freedom, dubbed hidden photon (sometimes paraphoton, dark photon, etc.)~\cite{Okun:1982xi}. Remarkably, such a new gauge boson can also be a suitable dark matter candidate~\cite{Nelson:2011sf, Arias:2012az,Graham:2015rva}. 
Therefore, this has become a popular test-case~\cite{Jaeckel:2010ni,Redondo:2010dp,Jaeckel:2013ija}.  
Small interactions between hidden photons and the 
Standard Model particles are most easily realized via kinetic mixing between the hidden photon and the ordinary photon~\cite{Holdom:1985ag,Foot:1991kb,Foot:1991bp}, which quite naturally arises in field theory via loop interactions of heavy messengers or similar effects in string theory. 
We will review the main properties of this type of interaction in Sect.~\ref{hidphot}.

Several experiments constrain the parameter space of hidden photons and many dedicated searches are running or planned for the future 
(see Refs.~\cite{Jaeckel:2013ija,Baker:2013zta,Essig:2013lka} for recent reviews). 
However, since both the mass ($m$) and the coupling ($\epsilon$) are a priori unknown, a wide parameter range needs to be explored and it is worthwhile to search for possible new tests. In particular in the mass range $m \sim {\rm meV}$, the limits are significantly weaker than in neighboring mass ranges.
Naively, low energy experiments probing this region should have a spatial size $\sim  1/m\sim {\rm mm}$.
Experiments probing the quantum mechanical interference of particles can be realized at this spatial size.
Famous examples are experiments testing the Aharanov-Bohm effect (ABE).
Experiments of the ABE type  have already been discussed as a possibility to search for a non-vanishing photon 
mass~\cite{Boulware:1989up}. 
In this paper we investigate the potential sensitivity of such experiments for hidden photons 
and consider further related experimental configurations that could improve the sensitivity.

The paper is structured as follows. In Sect.~\ref{hidphot} we briefly recall the essentials of hidden photons kinetically mixed with ordinary photons.
In Sect.~\ref{ABhid} we show how the ABE can be used to probe the existence 
of hidden photons and we give an estimate of the sensitivity of such experiments.
In Sect.~\ref{optim} we suggest improvements in the simple ABE setup.
We conclude in Sect.~\ref{conclusions}.

\section{Hidden photons}\label{hidphot}
The dynamics of hidden photons (HP) $a'_\mu$ in the interplay with visible photons $a_\mu$ can be
obtained from the Lagrangian
\begin{equation}
\label{Lagrange0}
{\mathcal{L}}=-\frac{1}{4} f_{\mu \nu} f^{\mu \nu}- \frac{1}{4} f'_{\mu \nu} f'^{\mu \nu}-
\frac{\sin \epsilon}{2} f'_{\mu \nu} f^{\mu \nu}+ \frac{\cos^2 \epsilon}{2} m^2 a'_\mu a'^\mu- j_\mu a^\mu,
\end{equation}
where
\be
f_{\mu \nu}= \partial_\mu a_\nu- \partial_\nu a_\mu \quad {\mbox{and}}\quad
f'_{\mu \nu}= \partial_\mu a'_\nu- \partial_\nu a'_\mu,
\ee
are the field strength tensors of the photon field and hidden photon fields.
The photon coupling to electromagnetic charges is implemented by minimal coupling to the electric four current $j^\mu$. 
We have included a mass term for the hidden photon, $m$, arising from a standard Higgs mechanism or a Stueckelberg mechanism.
The quantity $\epsilon$ accounts for the strength of the coupling between visible and hidden sectors and arises, e.g., at loop level via heavy messenger exchange.
It is constrained to be very small, typically in the range $10^{-12} \lesssim \epsilon \lesssim 10^{-3}$~\cite{Holdom:1985ag,Dienes:1996zr,Lukas:1999nh,Abel:2003ue,Blumenhagen:2005ga,Abel:2006qt,Abel:2008ai,Goodsell:2009pi,Goodsell:2009xc,Goodsell:2010ie,Heckman:2010fh,Bullimore:2010aj,Cicoli:2011yh,Goodsell:2011wn}, with quite some dependence.    

The kinetic mixing can be removed from the Lagrangian by rotating the fields to a new basis, 
with a massless photon-like field and a renormalised massive hidden photon field
\bea
A_\mu &=& a_\mu + a'_\mu  \sin \epsilon \label{a1},\\
 A'_\mu & =& a'_\mu \cos \epsilon \label{b1}. 
\eea
In this new basis the interaction between HPs and the electric current is evident, 
\bea
\partial_\nu F^{\mu \nu}& =& j^\mu \label{eqphoton},\\
\partial_\nu F'^{\mu \nu}+ m^2 A'^\mu & =&- \tan \epsilon \; j^\mu, \label{eqhidden}
\eea
where $  F^{\mu \nu}$ ($ F'^{\mu \nu}$) is now the field strength tensor of $A_\mu$ ($A'_\mu$).
The fact that in this basis both fields couple to the current $j^\mu$ will facilitate parts of our following analysis. 

Note that from now on, we will take $\tan\epsilon\sim \sin\epsilon\sim \epsilon$ given the smallness of $\epsilon$ required by experimental constraints. 
\section{Aharonov-Bohm effect for hidden photons}\label{ABhid}
The observable essence of the ABE is the path-dependent phase $\Phi$ of an electron wavefunction, 
which is shifted in the presence of an electromagnetic potential
\be
\exp \left(i e \oint_c \vec a \cdot d\vec x \right)\equiv \exp (i \Delta \varphi) .
\ee
Where the phase shift $\Delta \varphi$ is related to the magnetic flux enclosed by the path of the electron, $\Delta \varphi=e\Phi$. 
Conclusive experimental evidence for the ABE was obtained in 1986 with the experiment performed by Osakabe  et al \cite{Osakabe:1986zz}. They employed a toroidal magnet, surrounded by a superconducting  shielding to avoid magnetic leaking. An interference pattern was observed.

The theoretical modification of the ABE in presence of a non-zero photon mass was discussed in~\cite{Boulware:1989up}. 
Now let us highlight the analogous effect in the presence of hidden photons.
Consider eqs.~(\ref{eqphoton})-(\ref{eqhidden}) in the static limit,
\eqb\label{twoequations}
\nabla^2 {\vec{ A}}&=&- {\vec{j}},\\
\left(\nabla^2-m^2\right){\vec{ A'}}&=& \epsilon \,  {\vec{j}},
\eqf
where we have used the gauge condition $ A_0=\tilde A'_0=0$, which is consistent when $j^0=0$.  The equation of motion for the field ${\vec{ A}}$ is the usual equation for a massless gauge field, and the equation for the heavy HP is the Proca equation. 

In order to get an estimate of the potential sensitivity of an ABE applied to HP search, we consider the following idealized situation: 
a cylindrical solenoid, of radius $r_{sol}$, whose magnetic field (in the ordinary photon case) is entirely confined to the inside of the cylinder.
Using cylindrical coordinates and following~\cite{Boulware:1989up}, we find 
that the magnetic field associated to ${\vec{ A'}}$ given by ${\vec{B'}}=\nabla \times {\vec{A'}}$ is 
\bb
\label{Bprimemass}
{\vec{B'}}=-\hat z~ j \epsilon\,  \Theta(r_{sol}-r)-\hat z m^2 \epsilon\,  \Pi(r),
\ee
where $\Theta$ is the Heaviside step function,
$j$ is the current per unit height of the solenoid, of radius $r_{sol}$, and the function $\Pi(r)$ is given by
\begin{eqnarray}
\label{pieq}
\Pi(r)\!\!&=&\!\!-  j\bigg[ -\Theta(r-r_{sol}) K_0(mr) \int_0^{r_{sol}} r' dr' I_0(mr')\\\nonumber
&&+\Theta(r_{sol}-r)\left( K_0(mr)\int^r_0 r' dr' I_0(mr')+I_0(mr) \int_r^{r_{sol}} r'dr' K_0(mr')\right)\bigg].
\end{eqnarray}

Now we can return to the original basis 
and then find the conventional magnetic field (the one directly coupling to electric currents, cf. eq.~\eqref{Lagrange0}), 
given by ${\vec{b}}=\nabla \times {\vec{a}}$. From eq~(\ref{a1}) we have ${\vec{a}}={\vec{A}}-\epsilon{\vec{ A'}}$
and taking the curl of this equation we get
\bb
\vec b(r)= \hat z~ j \Theta(r_{sol}-r)+\hat z~j \epsilon^2\,  \Theta(r_{sol}-r)+\hat z~m^2 \epsilon^2\,   \Pi(r). \label{magneticb}
\ee
Therefore, the effect stops being purely topological in nature: 
there is an actual small leaking of magnetic field. 

To observe the ABE, an electron wave is split into two beams which are sent to pass by each side of the solenoid and then recombined after it. Besides the trivial dependence on the path, the interference pattern will depend on the magnetic flux enclosed in the cylinder. Varying the magnetic field while keeping the path of the electrons fixed,  the interference pattern will shift according to
\bb
\Delta \varphi = e\Phi,
\ee
where $\Phi$ is the flux enclosed by the solenoid. 

Without kinetic mixing, the flux is given by $\Phi_0= j\pi r_{sol}^2$. With mixing, the magnetic flux can be obtained by integrating Eq.~(\ref{magneticb}) over the surface enclosed by the path, and reads
\begin{eqnarray}
\Phi&=& \Phi_0\left(1+\epsilon^2\right) + m^2 \epsilon^2 \int_S   \Pi(r) dS
\\\nonumber
&=&\Phi_0\left(1+\epsilon^2\right) + m^2 \epsilon^2 \int^{R}_{r_{sol}}   \pi r  \Pi(r) dr,
\label{flux}
\end{eqnarray}
where we have chosen a circular path of radius $R$ around the solenoid.

The above formula is however not final yet, since it is not properly normalized in the limit  $m \rightarrow 0$. This is because the electric charge gets renormalized by the  photon-HP mixing~\cite{Jaeckel:2010xx, Endo:2012hp}. To properly renormalize and get a limit for the kinetic mixing one could follow \cite{Jaeckel:2010xx} (see \cite{Arias:2013isa} for details). 

To get an estimate of the sensitivity of an ABE let us briefly consider a solenoid of radius $r_{sol}$  with  internal magnetic field of $B=$ 1~T.  
Optimistically assuming that we can determine the internal magnetic field with a precision of $\Delta B= 10^{-8}~$T one quickly sees that we are sensitive only to $\epsilon> 10^{-4}$. 
Clearly the sensitivity of such experiment is very limited and can only probe a region in parameter space that has already been ruled out. The reason behind this poor sensitivity is that we are measuring a tiny signal on top of a huge standard model effect. In the following we aim to devise a null experiment where the standard model expectation is zero. 


\section{Two improved setups and two solution methods}\label{optim}

The major issue with the above setup is to overcome the limited sensitivity in the measurement of the phase shift of the electron beam. This obstacle can be evaded by directly measuring the leaking magnetic field with an ultralow-noise magnetometer like a 
Superconducting Quantum Interference Device
(SQUID) or using Nuclear Magnetic Resonance (NMR) techniques. To 
maximise the sensitivity one should modulate the B-field, and thus the signal, with a low frequency. 

Further, by shielding the solenoid with a layer of superconducting material we suppress effects of any imperfection of the solenoid, that could provide some B-field leakage of the solenoid. Even in this case we would still have a leakage through the shielding  because of the hidden photon effect. 
Thus, the setup would actually be a null experiment. 
 A related setup to look for hidden photons~\cite{Jaeckel:2008sz} along these lines was proposed some time ago. The main idea was to consider a superconducting shield near to a source of magnetic field.  
The derivation of the sensitivity of such an experiment was simplified to a one dimensional problem.

\begin{figure}[t]
\center
\includegraphics[scale=0.4]{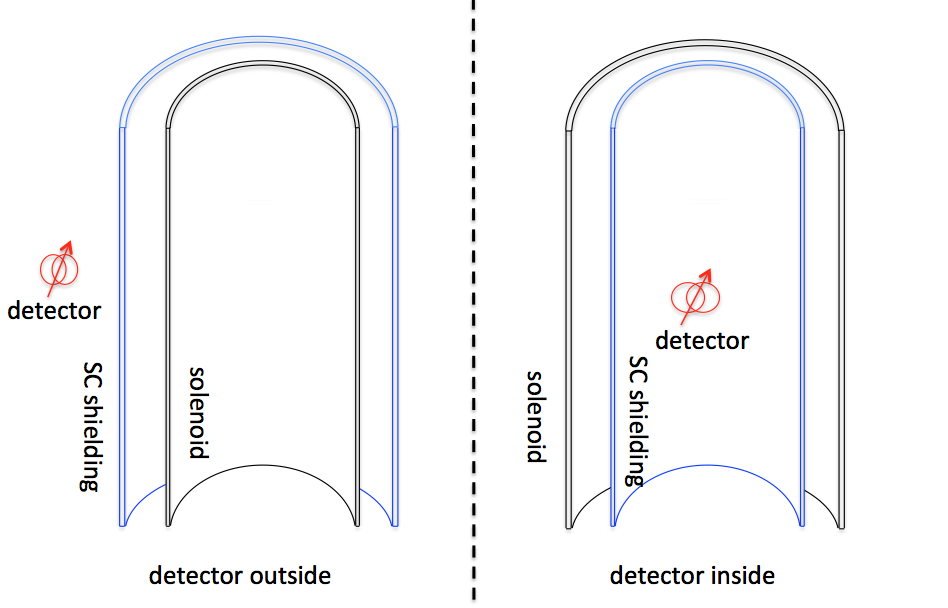}
\caption{Schematic configuration of the improved setups ${\bf Out}$ and ${\bf In}$.}
\label{fig:SetupsAB}
\end{figure}

Here we explore the potential of two setups, in a now realistic fully two-dimensional treatment. 
In the following we will treat these setups as if they were static. We assume that the B-field responds adiabatically to a sufficiently low frequency modulation of the current generating it. 
To implement the modulation we can ramp the current in the solenoid up and down. 
The two setups are: 
\begin{itemize}
\item[{\bf Out}] This setup considers a cylindric solenoid of radius $r_{sol}$, surrounded by a superconducting shielding of thickness $\delta$, (placed at a distance  $r_{sc}$ from the origin of coordinates) and a sensitive magnetometer outside the shielding.
\item[{\bf In}] This setup inverts the topology of first setup. It considers a cylindric solenoid of radius $r_{sol}$ containing a superconducting shielding of thickness $\delta$ and a sensitive magnetometer inside the shielding. 
\end{itemize}
Both configurations are sketched in figure \ref{fig:SetupsAB}. We choose to first cool down the superconducting shield and only then to switch on the current. In scenario {\bf Out} this is merely a convenient choice but in scenario {\bf In} this ensures that the magnetic field at the detector would vanish in the absence of HPs, and thus we have a true null experiment. 

We have solved the coupled equations for both setups {\bf Out} and {\bf In}.
However, since the direct solution of the two dimensional problem involves 
numerous boundary conditions and subtle numerical procedures at the limit of the machine precision,
we opted for double checking all results with a completely independent analytical approximation.
In the following subsections those two methods will be presented and compared.

\subsection{Method 1: Imposing boundary conditions}
Our first method for solving ${\bf Out}$ and ${\bf In}$ is straightforward: 
we solve the equations of motion in all spatial regions considering the corresponding electromagnetic currents, according to: solenoid, vacuum and shielding. The solenoid is treated as infinitely thin. In the superconductor, the current is given by the London current $\vec j_s=-\frac{M_L^2}2\,  a_\varphi(r) \, \hat  \varphi$ and hence proportional to the vector potential in the $\varphi$-direction. Thus, the equations are second order and linear in the fields in all four regions. The most general solution involves a combination of two functions with corresponding integration constants. These are determined by demanding that fields and derivatives are continuous. 

\subsubsection*{Setup ${\bf Out}$)} 
For this setup the superconducting shielding contains a cylindric solenoid of radius $r=r_{sol}$, surrounded with a current given by $\vec j=j \, \delta(r-r_{sol})\, \hat \varphi$, where $j$ is the current per unit height in the solenoid and $\varphi$ is the polar angle.
The detector (magnetometer) is placed outside the shielding. The four spatial regions are therefore:

\begin{itemize}
\item{\it{Region}}\, I: \, $0\leq r\leq r_{sol}$ (inside solenoid), $\vec j=j \, \delta(r-r_{sol})\,\hat \varphi$.
\item{\it{Region}} II:  $r_{sol}\leq r\leq r_{sc}$ (gap between the solenoid and the SC shielding).
\item{\it{Region}} III: $r_{sc}\leq r\leq r_{sc}+\delta $ (inside the superconductor), $\vec j_s=-\frac{M_L^2}2\,  a_\varphi(r) \, \hat  \varphi$.
\item{\it{Region}} IV: $r_{sc}+\delta\leq r$ (detector in vacuum), $\vec j=0$.
\end{itemize}
We normalize all scales by the distance from the center of the solenoid to the position of the magnetometer, so :   $r_i \rightarrow r_i/r_{det}$, $m_i \rightarrow m_i r_{det}$, $A_i \rightarrow A_i\, r_{det}$. 

\begin{figure}[h]
\centering
\includegraphics[width=0.7\textwidth]{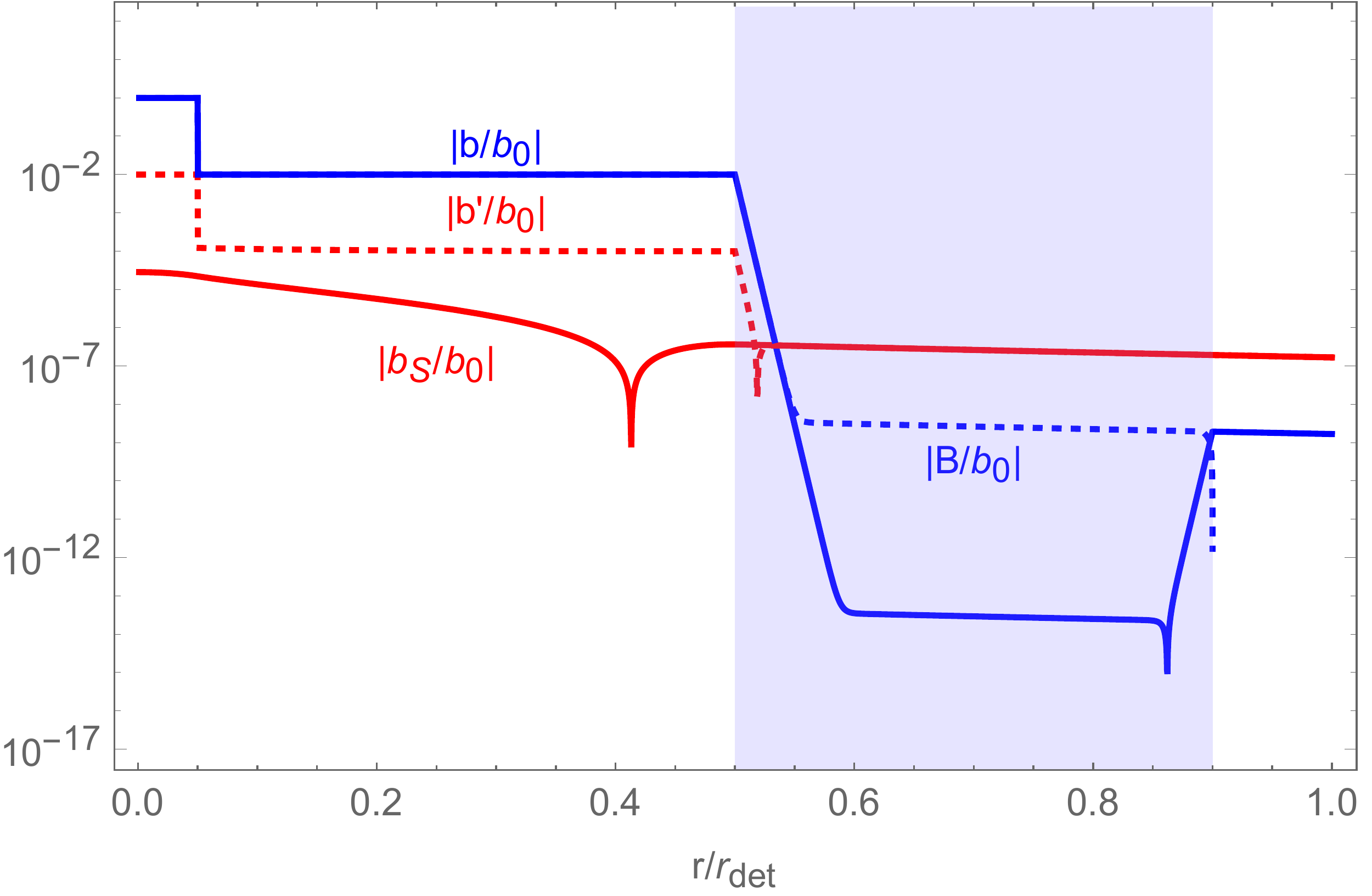}	
\caption{\footnotesize{Magnetic field ($b$, solid blue), photon-like propagation eigenstate B-field ($B$, dotted blue), sterile hidden photon state ($b_s$, solid red) and the hidden photon propagation state ($B'$, dotted red). The benchmark values used are $r_{sol}= 0.05\, r_{det}$,  $r_{sc}=0.5 \, r_{det}$, $\delta=0.4\, r_{det}$, $m=1/r_{det}$, $M_L= 300/r_{det}$ and $\epsilon=0.01$. }}
	\label{fig:fig_setB}
\end{figure}

The magnetic fields, normalized by the magnetic field of the solenoid, in each region are given by
\eqb
 B_{\rm I}(r)&=& (1+2 h_1),\,\,\,\,\, B'_{\rm I}(r)= \left(-\epsilon  m^2\Pi(r)/j+B'_{\rm hom1}(r)\right)\nonumber \\
B_{\rm II}(r)&=& 2 h_2,\,\,\,\,\,\,\,B'_{\rm II}(r)= \left(-\epsilon m^2\Pi(r)/j +B'_{\rm hom2}(r)\right)   \nonumber\\
B_{\rm III}(r)&=& \left( 1-\epsilon_{\rm eff}^2\right) \tilde{ \mathcal B_1}+\tilde{ \mathcal B_2} \epsilon_{\rm eff} ^2, \,\,\,\,\, B'_{\rm III}(r)=\epsilon_{\rm eff} \left(-\tilde{ \mathcal B_1}+\tilde{ \mathcal B_2}\right) \nonumber\\
B_{\rm IV}(r)&=&0,\,\,\,\,\, B'_{\rm IV}(r)=n_2 m\, K_0(mr).
\eqf
Here,
\eqb
\tilde{ \mathcal B}_i&=& m_i \left(p_i K_0(m_i\, r)- q_i I_0(m_i \, r) \right),\,\,\,\,\,\,\,\, i={1,2}, \,\,\,\,m_{1,2}=\left\{ M_L, m\right\}\\
\epsilon_{\rm eff}&=& \frac{M_L^2 \, \epsilon}{M_L^2-m^2},\\
B'_{\rm hom\, i}&=&m\left( - s_i K_0(m r)+ t_i I_0(m r) \right).
\eqf

\begin{figure}[h!]\centering
\includegraphics[width=0.7\textwidth]{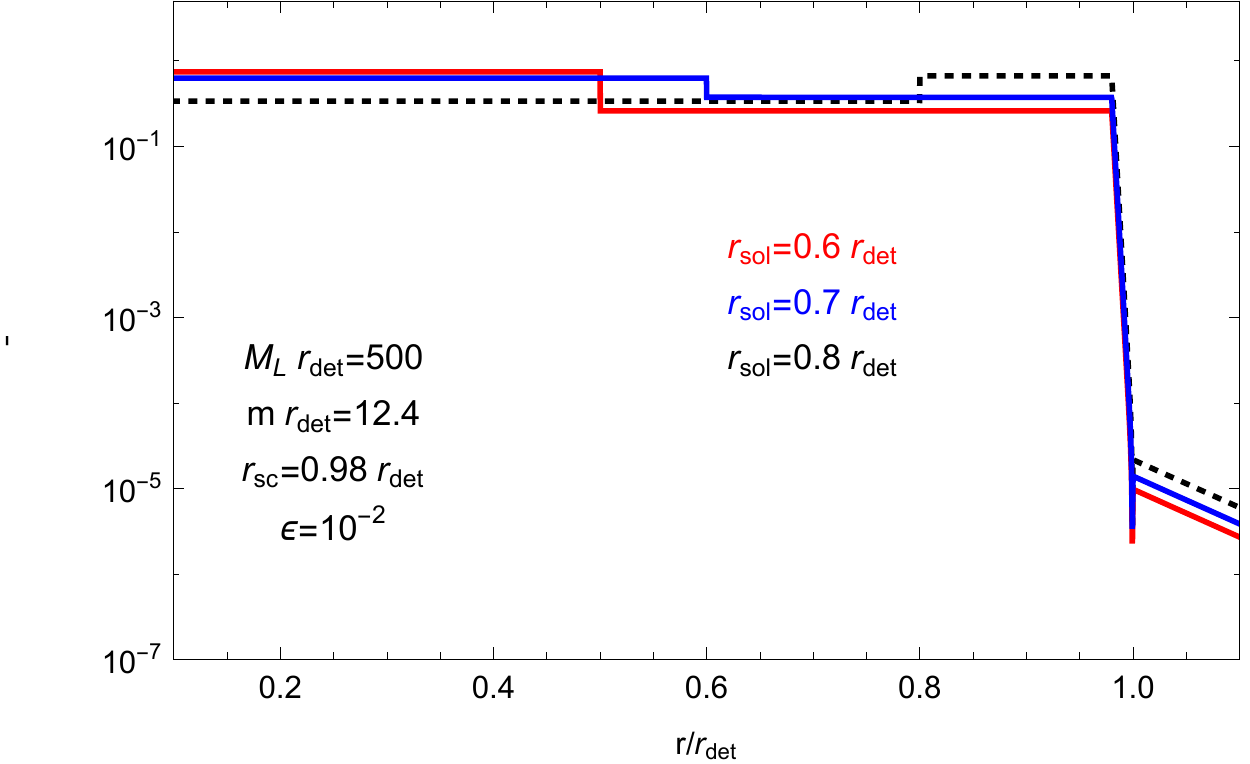}
	\caption{\footnotesize{Visible magnetic fields as a function of the distance for three different solenoid radius with a fixed position of the superconductor $r_{sc}=0.98 r_{det}$. All three configurations have $M_L r_{det}=500,\, m\,r_{det}=12.4,\, \delta =0.02\,r_{det}$ and $\epsilon=0.01$. The black dotted line is the optimal configuration for a solenoid radius of $r_{sol}=2~$cm and $m=10^{-4}$ eV.}}
\label{fig:fig_Bopt}
\end{figure}

Thus, there are 14 integration constants, $h_i, n_i, p_i,q_i, s_i, t_i, u_i$  ($ i={1,2}$), where $n_1$ and $u_i$ do not contribute to the magnetic fields, but appear in the vector potentials as $A_{\rm IV} (r)=n_1/r$ and $A_{\rm I}(r)=r/2+h_1 r+u_1/r$, $A_{\rm II} (r)=h_2 r+u_2/r$. By applying the continuity conditions we find $s_1=s_2=u_1=u_2=0$, $h_1=h_2$ and  $t_1=t_2$. The other constants have to be determined numerically.

In fig.~\ref{fig:fig_setB} we show the magnetic field as a function of the distance. The photon-like state (solid blue curve) and the sterile state (red solid curve) are produced by the solenoid, in the region $r\leq r_{sol}$.  Because of the external electromagnetic current there is a discontinuity in the visible magnetic field at $r=r_{sol}$. In the adjacent vacuum  region ($r_{sol}\leq r\leq r_{sc} $) the hidden photon state it is mainly given by $-\epsilon A$, since the sterile state is almost zero. Inside the superconductor (shaded region) the photon-like state is exponentially damped (Meissner effect). Meanwhile the sterile state propagates almost unperturbed (small suppression due to non-zero $m$). The photon-like state recovers 
at the end of the shielding due to the fact that the sterile state is not a propagation eigenstate. 


%
\subsubsection*{Setup ${\bf In}$)} 
In this setup, the detector is placed inside a superconducting shield, which in turn sits inside the cylindrical solenoid. Again, we distinguish four regions: 
\begin{itemize}
\item{\it{Region}}\, I: \, $0\leq r\leq r_{sc}$ (detector in vacuum), $\vec j=0$.
\item{\it{Region}} II:  $r_{sc} \leq r\leq r_{sc}+\delta$ (cylindrical superconducting shielding), $\vec j=j_s\,\hat \varphi$.
\item{\it{Region}} III: $r_{sc}+\delta \leq r\leq r_{sol}$ (vacuum between the superconductor and the solenoid), $\vec j=j \, \delta(r-r_{sol})\,\hat \varphi$.
\item{\it{Region}} IV: $r_{sol}<r$ (vacuum outside), $\vec j=0$. 
\end{itemize}
Here, $\vec j_s$ corresponds to the superconducting current, generated to expel the magnetic flux inside the shielding. 
The region IV carries no relevant information.
We will consider the ``zero field cooling'' case, where the superconducting sample is first cooled to its critical temperature at zero external magnetic field. In this case, the superconducting current can be written by the London current. Thus, $\vec j_s=-\left(M_L^2/2\right)\,  \vec a$, where $M_L^{-1} \propto$ to the penetration length of magnetic field in the superconducting sample, the so-called London mass. 
%

Defining new dimensionless variables,  parametrized by the cylinder radius, $r_i \rightarrow r_i/r_{sol}$, $m_i \rightarrow m_i r_{sol}$, $A_i \rightarrow A_i r_{sol}$, and 
solving for each region, we find the magnitude  of the magnetic fields (normalized by the magnetic field of the solenoid) for the propagation states as
\eqb
 B_{\rm I}(r)&=&  c_1,\,\,\,\,\, B'_{\rm I}(r)= c_2 m I_0 (m\, r)\nonumber\\
B_{\rm II}(r)&=& \left( 1-\epsilon_{\rm eff}^2\right) \mathcal B_1+\mathcal B_2 \epsilon_{\rm eff} ^2,\,\,\,\,\,\,\,B'_{\rm II}(r)= \epsilon_{\rm eff} \left(-\mathcal B_1+\mathcal B_2\right) \nonumber\\
B_{\rm III}(r)&=&1+g_1, \,\,\,\,\, B'_{\rm III}(r)= \left(B'_{\rm hom}-\epsilon m^2 \Pi(r)\right).
\label{eq_setA}
\eqf
Further,
\eqb
\mathcal B_i&=& m_i \left(d_i K_0(m_i\, r)- e_i I_0(m_i \, r) \right),\,\,\,\,\,\,\,\, i={1,2}, \,\,\,\,m_{1,2}=\left\{ M_L, m\right\}\\
\epsilon_{\rm eff}&=& \frac{M_L^2 \, \epsilon}{M_L^2-m^2},\\
B'_{\rm hom}&=&m\left(  f_2 I_0(m r)- f_1 K_0(m r) \right).
\eqf

\begin{figure}[t!]
\centering
\includegraphics[width=0.7\textwidth]{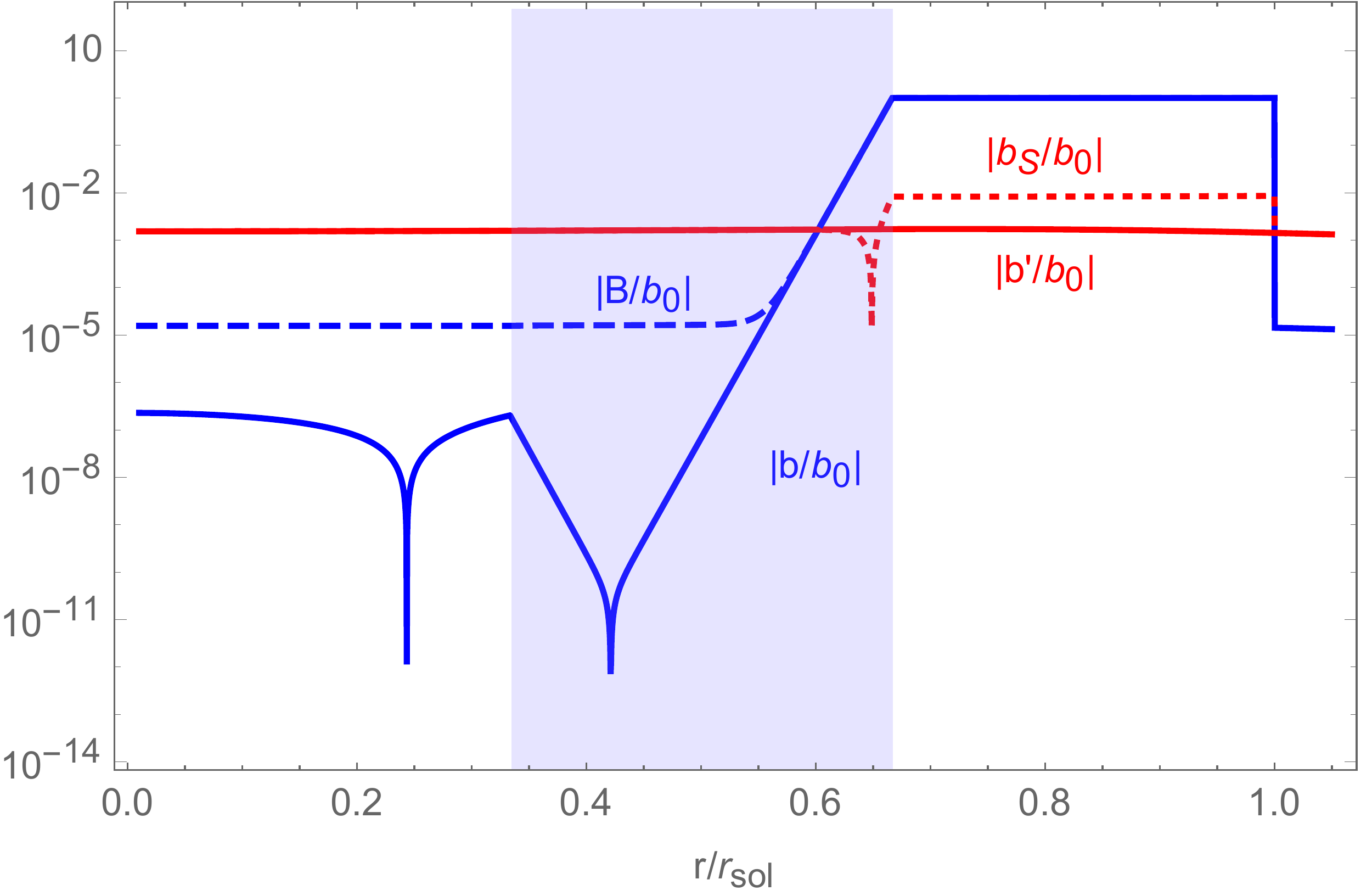}	
\caption{\footnotesize{Visible and hidden magnetic fields in different basis as a function of the dimensionless distance $r/r_{sol}$.  The blue solid and dashed lines correspond to the photon fields $b(r)$ and $B(r)$, respectively. The red solid and dotted curves correspond to the hidden fields $B'(r)+\epsilon B(r)$ and $b'(r)$ respectively. All magnetic fields are normalized by the magnetic field of the solenoid.  We have chosen the parameters $r_{sc}=1 /3\,r_{sol},\,  \delta=1/3\, r_{sol}$, $m=1/r_{sol}$, $M_L=100/r_{sol}$ and $\epsilon=0.01$.}}
	\label{fig:fig_setA}
\end{figure}	

There are in total ten integration constants $c_i, d_i, e_i,f_i, g_1$ ($ i={1,2}$), and one more that does not contribute to the magnetic field, but appears in the vector potential $A_{\rm III}$(r) and goes as $g_2/r$ (this last one is of course the key for the original ABE). The function $\Pi(r)$ is the same as in eq.~(\ref{pieq}). 
Imposing continuity to the vector potentials ($A$ and $A'$) and their derivatives in each boundary of region I, II and III, we find $g_1=f_2=0$. It is hard to find analytical expressions for the rest of the integration constants, so we again have computed them numerically. In fig.~\ref{fig:fig_setA} we show the magnetic fields normalized by the magnetic field of the solenoid as a function of the dimensionless distance $r/r_{sol}$.  The solenoid  mainly produces the photon-like state $B-\epsilon B'$ (solid blue line) and a small sterile component $b_S=B'+\epsilon B$ (solid red line). Since the photon-like state couples directly to the electric current, 
it gets exponentially damped inside the superconductor because of the Meissner effect. On the other hand the sterile component can penetrate unperturbed through the solenoid.  Because of this component, the photon-like state reappears inside the superconducting shielding.  The blue dashed and red dotted lines correspond to the  $B(r)$ and $B'(r)$ fields, respectively. For better representation we have chosen $r_{sc}=r_{sol}/3,\,  \delta= r_{sol}/3, \, m r_{sol}=1,\, M_L r_{sol}=100$ and $\epsilon =0.01$.

A notable feature of fig.~\ref{fig:fig_setA} (see also fig.~\ref{fig:fig_Aopt} below) is the change of sign in the magnetic field $b$ inside the shielding that is visible as a sharp dip in the logarithmic plot. We can understand this from the requirement that the total magnetic flux in that region is zero, inherited from the periodicity of the wave-function of Cooper pairs inside the superconductor $0=\oint d\vec l \cdot \vec a = \int_S d\vec s \cdot \vec B$. 

In a realistic setup, the London mass ($M_L$) is typically of the order of the eV, thus  much  bigger than the mass of the hidden photon we are interested in probing. It seems useful then to take the limit  $M_L\rightarrow \infty$ in the equations of the magnetic fields inside the superconductor. Looking at eqs.~(\ref{eq_setA}) we can see that $K_0(M_L r) \rightarrow 0$ in this limit. In fig.~\ref{fig:fig_Aopt} we plotted the visible magnetic field $b(r)= B-\epsilon B'$ for three different configurations of the shielding in the realistic scenario that $M_L \delta \gg 1$. For plotting we have  fixed the thickness of the  superconductor to $\delta = 0.05 \, r_{sol}$, $M_L r_{sol}=500$ and $m\,r_{sol}=10$, but we vary the position of the superconductor with respect to the solenoid. For a given radius of the solenoid, the strength of the magnetic field in the region of the detector can be optimized by choosing the appropriate position of the superconducting shielding (black dashed line). To do so we have used the analytical approximation we describe in the next subsection (see eq.~(\ref{btotin})).

\begin{figure}[t]
\centering
\includegraphics[width=0.7\textwidth]{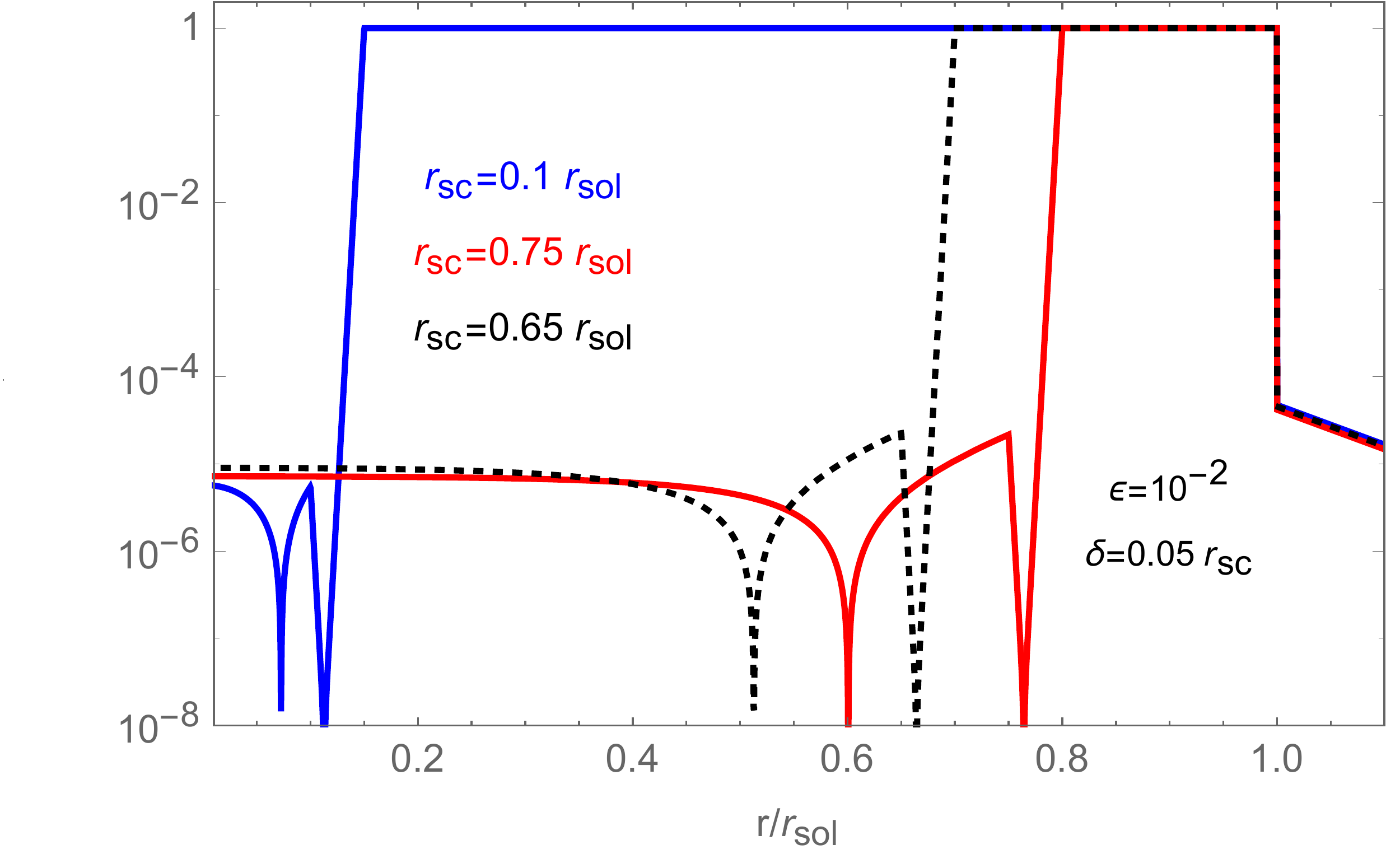}
	\caption{\footnotesize{Visible magnetic field (normalized by the input magnetic field) as a function of $r/r_{sol}$ for different configurations.
	 }}
\label{fig:fig_Aopt}
\end{figure}


\subsection{Method 2: Analytical approximation and comparison}

One can achieve a substantial simplification of the problem
by assuming that the London penetration length is
much smaller than the other length scales involved in 
the problem.
This is a realistic approximation as long as the hidden photon mass is not too large, 
\be
m\ll M_L\, |\log \epsilon^2| .  
\ee
The log arises because for a null experiment we want to sufficiently suppress the standard leaking of the B-field, $\sim B\exp(-M_L \delta)$, such that it is smaller than the regenerated field from the hidden photon $\sim \epsilon^2\exp(-m\delta)$. 

Under this assumption the electrical current
in the superconductor can be taken as a mere surface current $j_{s}$.
Thus, the description of the whole system will be given by a
 solenoid with radius $r_{sol}$ with current per unit height $j$,    
 and a superconductor with radius $r_{sc}$ with current $- j_{s}$.
In a magnetostatic setting one can solve the equations in
the propagation eigenstate basis
\bea
-\nabla^2 \vec A&=& \vec j, \\
 \vec \nabla \vec A&=&0\quad,
\eea
and
\bea
(-\nabla^2+m^2) \vec A'&=&\epsilon \, \vec j, \\
\vec \nabla \vec A'&=&0, 
\eea
where $\vec B= \vec \nabla \times \vec A$ and
$\vec B'= \vec \nabla \times \vec A'$.
For a single solenoid with current density $j$
the massless photon state has simply
\be
\vec B = \hat z   j\theta(r_{sol}-r).
\ee
The magnetic field for a single solenoid with current density $\epsilon j$ and a photon with mass $m$ is given by eq.~\eqref{Bprimemass}.

The same solutions hold for the superconductor, 
only that the currents are replaced by $j \rightarrow -j_{s}$.
The propagation eigenstates don't mix and thus, 
the total fields are simply the superpositions 
of the fields produced by the solenoid and
the superconductor
\bea
\vec B_{tot}&=&B_{sol}+B_{SC}, \\
\vec B'_{tot}&=&B'_{sol}+B'_{SC}.
\eea
Of course the problem is not completely solved yet since
the superconducting current $j_{s}$ is not a free parameter of the experiment,
it has to be determined in terms of the initial current $j$, the model parameters
$\epsilon$, $m$,
and the geometric configuration of $r_{sol}$, and $r_{sc}$.
This can be achieved by using the physical condition that
the superconductor will readjust its surface current to maintain zero inner magnetic flux 
\be\label{scCond2}
\Phi_{b}= 2 \pi \int_0^{r_{sc}} r  b_{tot}(r)=0 ,
\ee
where $b(r)$ is the interaction eigenstate
\be
\left(
\begin{array}{c}
b\\
b'
\end{array}\right)=
\left(
\begin{array}{cc}
1 & -\epsilon\\
\epsilon & 1
\end{array}\right)
\left(
\begin{array}{c}
B_{tot}\\
B'_{tot}
\end{array}\right).
\ee
Solving the condition (\ref{scCond2}) for the scenario {\bf Out} gives to leading order
\be\label{jSCout}
j_{s}|_{\bf Out}=\left(\frac{r_{sol} }{r_{sc}}\right)^2j+{\mathcal{O}}(\epsilon^2).
\ee
Whereas for the scenario ${\bf In}$ one finds
\be\label{jSCin}
j_{s}|_{\bf In}=j+\epsilon^2 j
\left[
r_{sol} K_1(m r_{sol} )-r_{sc} K_1(m r_{sc})
\right]
F^1_{0,1}\left(2,m^2 r_{sc}^2 /4\right)
+{\mathcal{O}}(\epsilon^4)
.
\ee 
Those currents have to be replaced
in the final expression for $b(r)$.
After some simplifications
and approximations to leading order in $\epsilon$, the magnetic fields
on the detector side of the superconducting surface
turn out to be rather simple. 
In scenario ${\bf Out}$ one finds
\be\label{btotout}
b_{tot}(r>r_{sc}>r_{sol})=\epsilon^2 \frac{r_{sol}}{r_{sc}} m j\left\{
 r_{sol} I_1(mr_{sc})-r_{sc} I_1(m r_{sol}))
\right\}K_0(mr)+{\mathcal{O}}(\epsilon^4),
\ee
and  in scenario ${\bf In}$ one finds
\bea
\label{btotin}
&&\!\!\!\!\!\!\!\!\!\!\!\!\!\!\!\!\!\!\!\!\! b_{tot}(r<r_{sc}<r_{sol}) = \\ \nonumber
&&
\epsilon^2 \frac{j }{r_{sc}}
\left(
r_{sc}m I_0(mr)-2I_1(mr_{sc})
\right)
\left(
r_{sol} K_1(m r_{sol})-r_{sc}K_1(m r_{sc})
\right)+{\mathcal{O}}(\epsilon^4).
\eea
In both scenarios ${\bf Out}$ and ${\bf In}$ one verifies the limits
\be
\lim_{m\rightarrow 0} b_{tot}|_{det}=\lim_{m\rightarrow \infty} b_{tot}|_{det}=0,
\ee
which implies  that no charge renormalization is necessary.

\begin{figure}[t]
\centering
\includegraphics[width=0.7\textwidth]{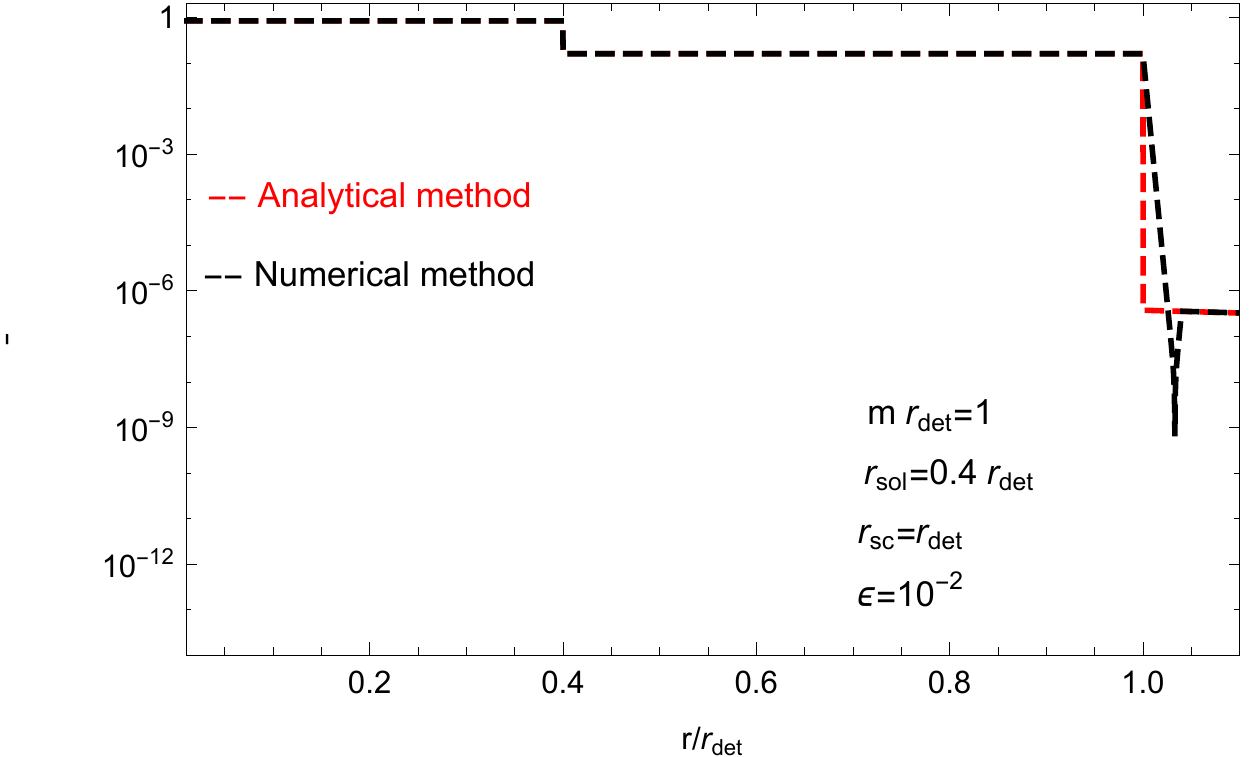}
\caption{\footnotesize Visible magnetic fields $b(r)$ for scenario ${\bf Out}$.
The black line is the numerical solution solution for a superconducting shielding
with finite thickness and finite London mass ($M_L r_{det}=500$).
The red line is the analytical approximation which gives in the outside region ($r/r_{det}\ge1$)
the magnetic field strength (\ref{btotout}).}
\label{fig:compout}
\end{figure}

\begin{figure}[t]
\centering
\includegraphics[width=0.7\textwidth]{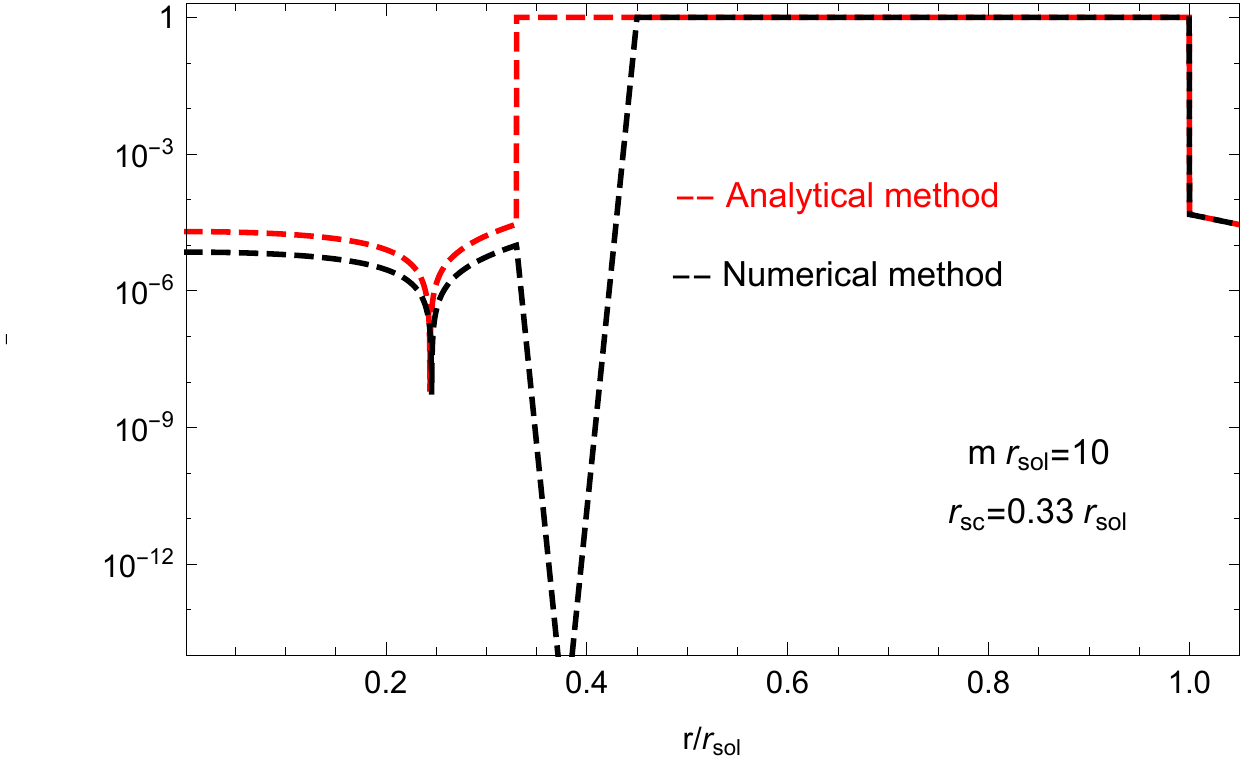}
\caption{\footnotesize Visible magnetic fields $b(r)$ for scenario ${\bf In}$.
The black line is the numerical solution solution for a superconducting shielding
with finite thickness and finite London mass ($M_L r_{sol}=500$).
The red line is the analytical approximation which gives in the inside region ($r/r_{sol}\le 0.2$) 
the magnetic field strength (\ref{btotin}).}
\label{fig:compin}
\end{figure}

In order to check the robustness of those analytical results,
one compares the approximation from this subsection to the full numerical solution obtained
in the previous subsection.
In figs. \ref{fig:compout} and \ref{fig:compin} a comparison
of the numerical solution and the analytical approximation is shown for both scenarios ${\bf Out}$ and ${\bf In}$. 
In the region of interest (location of the detector) one can see that both methods are in very good agreement.%

\subsection{Experimental reach}

\begin{figure}[hbt]
\center
\includegraphics[scale=0.9]{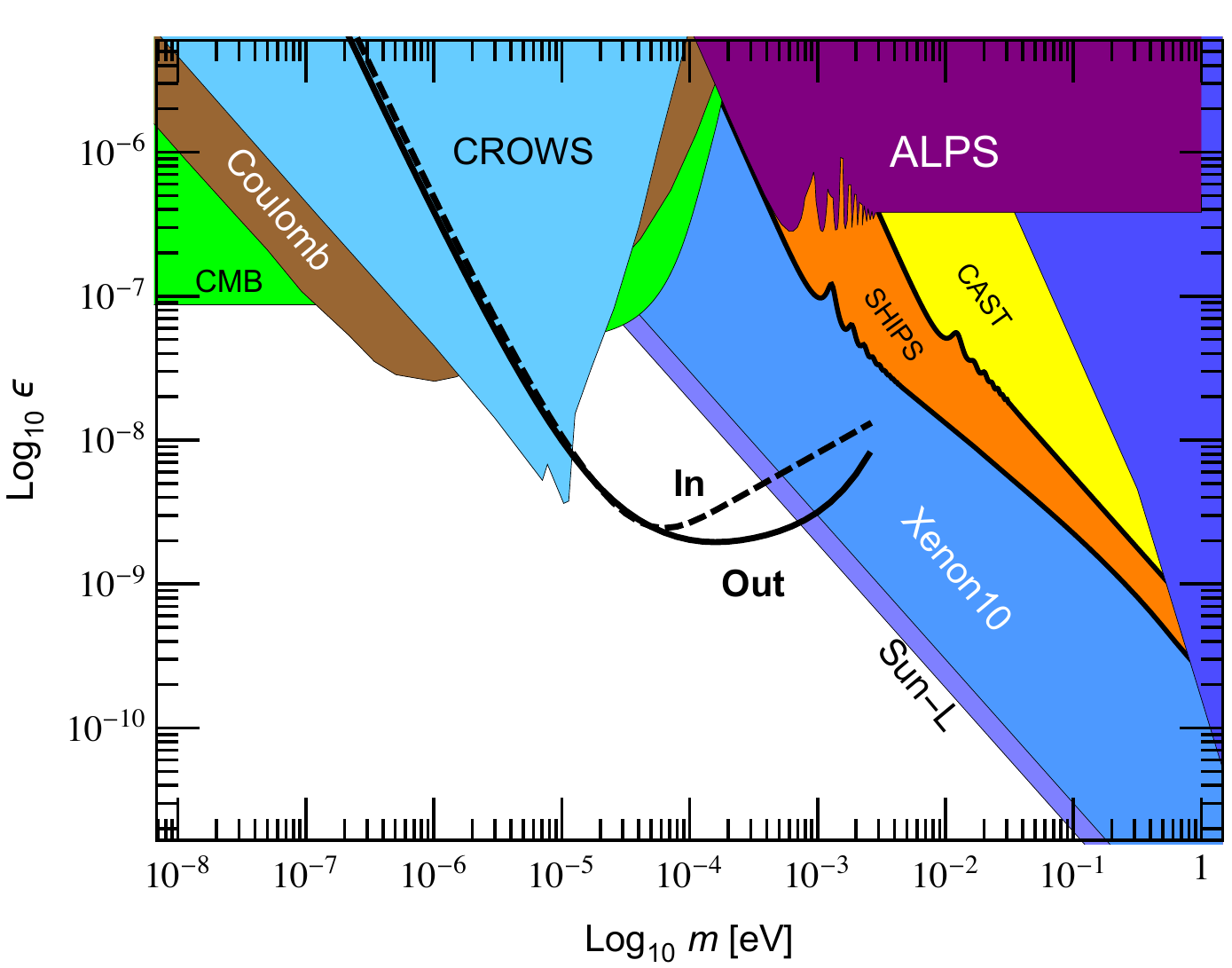}
\caption{\small Sensitivity range of setups ${{\bf Out}}$ and ${{\bf In}}$ with detector outside or inside a solenoid with radius $r_{sol}=2$ cm and current $j=1$ T and detector sensitivity $b_{det}=10^{-18}$ T. We assume that the superconducting shielding cancels all standard leaking of the B-field, i.e. $M_L\to \infty$. 
In scenario ${\bf Out}$ (solid line) the superconducting shielding is placed at $r_{sc}=2.4~{\rm cm}$ and the detector at $r=1.01\, r_{sol}$. 
For scenario ${\bf In}$ (dashed line) we place the shield at $r_{sc}=1.3~{\rm cm}$ and the detector at $r=0$. 
Colored areas are regions excluded by experiments and astrophysical observations (compilation adapted from~\cite{Jaeckel:2010ni,Jaeckel:2013ija,An:2013yua,Redondo:2013lna,An:2013yfc,Betz:2013dza,Graham:2014sha,Schwarz:2015lqa,Vinyoles:2015aba}).
}
\label{fig:hp_bound2}
\end{figure}

Given the very good agreement between the two methods
of calculating the magnetic fields, 
the experimental sensitivity analysis
will be based on the simpler analytical approximation.
In the following we assume that the detector has a sensitivity 
to magnetic fields of order $b_{det}=1\cdot 10^{-18}$~T. 
These sensitivities are within reach of the most precise magnetometers for integration times of order of one week, see for instance~\cite{Vasilakis:2008yn,Tullney:2013wqa}. 
Thus, deviations from the classical null result,  like the hidden photon induced $b_{tot}$, 
would be detectable if
\be b_{tot}\ge b_{det}.
\ee
Inserting  (\ref{btotout}) or (\ref{btotin}) into this inequality
and solving for $\epsilon$ one gets the expected experimental
sensitivity range as a function of 
$j,\,b_{det},\,m,\,r_{sol},\,r_{sc}$, and $r$.
For the scenario ${\bf Out}$ one gets
\be\label{boundout}
\epsilon_{A}\ge 
\sqrt{
b_{det}\frac{r_{sc}}{j m r_{sol}} }
\frac{1}{\sqrt{\left[
 r_{sol} I_1(r_{sc}m)-r_{sc} I_1(r_{sol} m)
\right]K_0(mr)}},
\ee
and for the scenario ${\bf In}$ one gets
\be\label{boundin}
\epsilon_B= \sqrt{\frac{b_{det} r_{sc}}{j }}
\frac{1}{\sqrt{\left(
r_{sc}m I_0(mr)-2I_1(r_{sc}m)
\right)
\left(
r_{sol} K_1(m r_{sol})-r_{sc}K_1(m r_{sc})
\right)}}.
\ee
For the analysis of the parameter range the radius
of the solenoid was fixed to $r_{sol}=2$~cm and the electrical
current density was taken to be $j=1$~T.
Also the radial position of the detector $r$ was fixed for 
both experimental scenarios.
For the scenario ${\bf Out}$ the position of the detector
was chosen to be $r=1.01\,r_{sol} $ (close to the superconducting shielding).
While for the scenario ${\bf In}$ the position of the detector
was chosen to be at the center of the solenoid $r=0$.
Those settings left
$r_{sc}$ and $m$ as free parameters of (\ref{boundout}) or (\ref{boundin}).
Since other experiments leave an interesting
mass window at $m\sim 10^{-4}$~eV, cf. Fig.~\ref{fig:hp_bound2}, it would be interesting to
choose an optimal position for the superconducting shielding $r_{sc}$
for this mass range.
This is achieved by setting $m\sim 10^{-4}$~eV and maximizing
$b_{tot}$ as a function of $r_{sc}$.
For the scenario ${\bf Out}$ one finds good sensitivity for  
\be
r_{sc}|_{{\bf Out}}=2.4\,{\rm cm}
\ee
while for the scenario ${\bf In}$ 
\be
r_{sc}|_{{\bf In}}=1.3\,{\rm cm}.
\ee
turns out to be a good choice. 

Now one can take those tuned experimental choices of the 
superconductor radii $r_{sc}|_{\bf Out}$ and $r_{sc}|_{\bf In}$ and insert them in 
eqs. (\ref{boundout}) and (\ref{boundin}), respectively, while allowing
$m$ to take arbitrary values. This will finally give the experimental sensitivity
range as a function of the hidden photon mass $\epsilon=\epsilon(m)$.
In fig.~\ref{fig:hp_bound2} the sensitivity range of
(\ref{boundout}) and (\ref{boundin}) is compared to 
the currently existing bounds~\cite{Jaeckel:2010ni,Jaeckel:2013ija}, see also~\cite{An:2013yua,Redondo:2013lna,An:2013yfc,Betz:2013dza,Graham:2014sha,Schwarz:2015lqa,Vinyoles:2015aba}.
In the mass range of $10^{-4.5}$ eV$<m<10^{-1}$ eV 
one finds a sensitivity range which would be superior to existing
limits.
Further, one can check to which extend
this high sensitivity is sensible to the ``optimal'' choice of $r_{sc}$.
It turns out this dependence is rather mild,
for example when varying $r_{sc}$ in scenario ${\bf In}$
by 50\% only varies the maximal sensibility by 50\%, which 
would appear as small effect on the logarithmic scale of Fig.~\ref{fig:hp_bound2}.

\section{Conclusions}
\label{conclusions}
In this paper we studied the potential of using quantum interference experiments
with cylindrical symmetry for exploring the parameter space of
models with hidden photons ($\epsilon$, $m$).
As starting point we estimate the reach for a classical Aharonov-Bohm type phase measurement. 
It is found that in this case, the systematic and experimental uncertainties prevent 
a competitive experimental reach.
We suggest an improvement of the experimental setting such that it becomes
a null experiment in the spirit of \cite{Jaeckel:2008sz}.
For this improvement we study two experimental scenarios differing in the position of the detector, inside and outside the solenoid. Predictions for both scenarios are obtained by two independent
methods, one partially relying on numerical methods and the other one
using an analytical approximation.
For both scenarios a very good agreement between the two methods is achieved.
Based on the analytical approximation, we calculate
the discovery potential of the two improved scenarios in comparison
to existing experimental and observational bounds. We see that a significant area of un-probed parameter space can be tested with experiments of this type.  

\section*{Acknowledgements}
 P.~A. and J.~J. would like to thank to the University of Zaragoza for hospitality. P.~A. acknowledges  support from Fondecyt  project 11121403 and ANILLO Atlas Andino, ACT 1102. 
J.~J. gratefully acknowledges support from the TransRegio TR33 ``The Dark Universe''.
The work of B.~K. and C.~D. was supported by proj.\ Fondecyt 1120360 
and ANILLO Atlas Andino, ACT 1102.
The work of M.~D.\ was supported by Fondecyt 1141190. and  ANILLO Atlas Andino, ACT 1102.
The work of J.~R. was supported by the Ramon y Cajal Fellowship 2012-10597 from the Spanish Ministry of Economy and Competitivity.




\newpage
%

\end{document}